\documentclass[english,oneside]{elsart}
\usepackage[T1]{fontenc}
\usepackage[latin9]{inputenc}
\usepackage{amsbsy}
\usepackage{amstext}
\usepackage{graphicx}

\makeatletter

\providecommand{\tabularnewline}{\\}

\usepackage{amsbsy}

\usepackage{amstext}



\makeatother

\usepackage{babel}

\begin{document}
\begin{frontmatter}

\begin{frontmatter}

\title{A general purpose Fortran 90 electronic structure program for conjugated
systems using Pariser-Parr-Pople model}

\author{Priya Sony$^{1}$, Alok Shukla$^{2}$}

\address{Department of Physics, Indian Institute of Technology, Bombay, Powai,
Mumbai 400076, INDIA}

\thanks{e-mail: psony@phy.iitb.ac.in}

\thanks{Author to whom all the correspondence should be addressed. e-mail: shukla@phy.iitb.ac.in}
\begin{abstract}
Pariser-Parr-Pople (P-P-P) model Hamiltonian has been used extensively
over the years to perform calculations of electronic structure and
optical properties of $\pi$-conjugated systems successfully. In spite
of tremendous successes of \emph{ab initio} theory of electronic structure
of large systems, the P-P-P model continues to be a popular one because
of a recent resurgence in interest in the physics of $\pi$-conjugated
polymers, fullerenes and other carbon based materials. In this paper,
we describe a Fortran 90 computer program developed by us, which uses
P-P-P model Hamiltonian to not only solve Hartree-Fock (HF) equation
for closed- and open-shell systems, but also for performing correlation
calculations at the level of single configuration interactions (SCI)
for molecular systems. Moreover, the code is capable of computing
linear optical absorption spectrum at various levels, such as, tight
binding (TB) Hückel model, HF, SCI, and also of calculating the band
structure using the Hückel model. The code also allows the user to
solve the HF equation in the presence of finite external electric
field, thus, permitting calculations of quantities such as static
polarizabilities and electro-absorption spectra. Additionally, it
can perform transformation of P-P-P model Hamiltonian from the atomic
orbital (AO) representation (also called site representation) to the
molecular orbital (MO) one, so that the transformed matrix elements
can be used for high level post-HF calculations, such as, full CI
(FCI), quadruple CI (QCI), and multi-reference singles-doubles CI
(MRSDCI). We demonstrate the capabilities of our code by performing
calculations of various properties on conjugated systems such as $trans$-polyacetylene
($t$-PA), poly-\emph{para}-phenylene (PPP), poly-\emph{para}-phenylene-vinylene
(PPV), \textit{oligo}-acenes, and graphene nanodisks. \end{abstract}
\end{frontmatter}
\end{frontmatter} \textbf{Program Summary} \\
 \emph{Title of program:} ppp.x \\
 \emph{Catalogue Identifier:} \\
 \emph{Program summary URL:} \\
 \emph{Program obtainable from:} CPC Program Library, Queen's University
of Belfast, N. Ireland \\
 \emph{Distribution format:} tar.gz\\
 \emph{Computers :} PC's/Linux\\
 \emph{Linux Distribution:} Code was developed and tested on various
recent versions of Fedora including Fedora 11 (kernel version 2.6.29.4-167)\\
 \emph{Programming language used:} Fortran 90\\
 \emph{Compilers used:} Program has been tested with Intel Fortran
Compiler (non-commercial version 11.1) and gfortran compiler (gcc
version 4.4.0) with optimization option -O. \\
 \emph{Libraries needed:} This program needs to link with LAPACK/BLAS
libraries compiled with the same compiler as the program. For the
Intel Fortran Compiler we used the ACML library version 4.3.0, while
for gfortran compiler we used the libraries supplied with the Fedora
distribution.\\
 \emph{Number of bytes in distributed program, including test data,
etc.:} size of the tar file ...... bytes\\
 \emph{Number of lines in distributed program, including test data,
etc.:} lines in the tar file .......\\
 \emph{Card punching code:} ASCII\\
 \emph{Nature of physical problem:} The problem of interest at
hand is the electronic structure of $\pi$-conjugated systems. For
such systems, the effective $\pi$-electron P-P-P semi-empirical model
Hamiltonian proposed by Pariser, Parr, and Pople offers an attractive
alternative as compared to the \emph{ab initio} approaches. The present
program can solve the HF equations for both open- and closed-shell
systems within the P-P-P model. Moreover, it can also include electron
correlation effects at the singles CI level. Along with the wave functions
and energies, various properties such as linear absorption spectra
can also be computed. \\
 \emph{Method of Solution:} The single-particle HF orbitals of
a $\pi$-conjugated system are expressed as linear combinations of
the $p_{z}$-orbitals of individual atoms (assuming that the system
is in the $xy$-plane). Then using the hopping and Coulomb parameters
prescribed for the P-P-P method, the HF integro-differential equations
are transformed into a matrix eigenvalue problem. Thereby, its solutions
are obtained in a self-consistent manner, using the iterative diagonalizing
technique. The HF orbitals thus obtained can be used to perform a
variety of calculations such as the SCI, linear optical absorption
spectrum, polarizabilty, electro-absorption spectrum, \textit{etc}.\\
 \emph{Running Time:} The examples provided each only take a few
seconds to run. For a large molecule or a polymer, however, the run
time may be a few minutes.\\
 \emph{Unusual features of the program:} None

\section{Introduction}

The studies of electronic structure and optical properties of $\pi$-conjugated
systems have fascinated physicists and chemists alike for a long time\cite{barford-book,salem},
because of their possible applications in opto-electronic devices
such as OLEDs, OFETs, organic lasers, solar cells, \textit{etc}\cite{polymer review}.
Ever since graphene and its nano-structures such as graphene nano-ribbons,
nano-disks, \textit{etc}. have been synthesized\cite{geim-nature},
interest in the study of $\pi$-conjugated systems has multiplied
many folds\cite{geim-rmp}.

In the 1950's, Pariser, Parr, and Pople proposed a simple model which
incorporates the essential physics of correlated conjugated systems
in a very elegant manner\cite{PPP}. In this model only $\pi$-electrons
are considered explicitly, while the effect of $\sigma$-electrons
is included in an implicit manner in terms of various parameters.
Moreover, long range electron-electron interactions are taken into
account by means of suitable Coulomb parameters. Thus, one reduces
the number of degrees of freedom involved in the underlying Schrödinger
equation considerably. Because of the lack of large scale computational
facilities during the 1950's such an approach was unavoidable even
for relatively small molecules such as benzene. However, the remarkable
fact is that in spite of so many approximations involved, P-P-P model
based calculations were extremely successful in describing the electronic
structure, in general, and the optical properties of such systems,
in particular\cite{salem}.

Now, the question arises that in the present times, when computational
power has grown by orders of magnitude as compared to those in 1950's,
why should one be still interested in using the P-P-P model, particularly,
when so many sophisticated \textit{ab initio} electronic structure
programs are available. In our opinion there are two main reasons
behind the continued use of the P-P-P model at present: (a) One gets
to understand the electronic and optical properties of such systems
within a rather simplified picture, in terms of the dynamics of the
$\pi$-electrons, (b) Ever since the revival of interest in the $\pi$-conjugated
polymers because of their possible applications in opto-electronics,
the size of the systems being investigated these days is becoming
quite large consisting of hundreds of atoms, pendant side groups,
and terminal functionalities. Therefore, simultaneous high quality
\textit{ab initio} description of both the ground and the excited
states of such systems (needed to describe their linear and nonlinear
optical properties) is virtually impossible. While, the P-P-P model,
with its reduced degrees of freedom, allows one to compute low-lying
excited states of rather large systems using the present computational
capacities.

Indeed a large number of calculations on various conjugated systems
have been performed in last several years using the P-P-P model Hamiltonian
with a great deal of success\cite{barford-book,sumit-review,jug-ppp,ppp-soos,ppp-bredas,ppp-ramasesha,ppp-barford,ppp-sumit}.
In our own group and with our collaborators, we have performed several
calculations on systems as diverse as phenyl disubstituted polyacetylenes
(PDPAs)\cite{shuklamazumdar83,shuklaPRB-62,shukla-synthmet116,shuklaCP-300,shuklaPRB-69,sony-pdpa,priya-pdpa-synth},
PPVs\cite{shukla-ppv-prb65,shuklaPRB-67,shukla-ppv-synth}, and \textit{oligo}-acenes\cite{priya-acene-synth,priya-acene-prb,priya-acene-jcp}.
In all the afore-mentioned works the HF calculations and the transformation
of the Hamiltonian from the AO to the MO representation were performed
using a Fortran 77 code, developed earlier in our group\cite{shuklaF77}.
However, the code in question lacked the capability of performing
unrestricted Hartree-Fock (UHF) calculations, and thus, was limited
only to closed-shell systems. Moreover, Fortran 77 language did not
allow dynamic memory allocation, resulting in a dependence on compile-time
dimensional parameters. Therefore, we have rewritten the code in a
modern programming language, namely, Fortran 90, which allows it to
utilize dynamic memory allocation, thereby freeing it from various
array limits, and resultant artificial restrictions on the size of
the molecules. Thus our program can be used on a given computer until
all its available memory is exhausted. The present computer program
can perform restricted Hartree-Fock (RHF) calculations on closed-shell
systems, as well as unrestricted-Hartree-Fock (UHF) calculations on
open-shell systems. Moreover, like the earlier version of the code\cite{shuklaF77},
it can also carry out correlation calculations at the level of singles
configuration interactions (SCI) for molecular systems. Additionally,
the code is capable of computing optical spectrum at various levels,
such as, TB, HF, and SCI, and also of calculating the band structure
using Hückel model. The program also allows user to solve HF equation
in the presence of an external electric field, thus, permitting calculation
of dielectric polarizabilities and electro-absorption spectrum of
conjugated molecules. Additionally, it can perform transformation
of P-P-P model Hamiltonian from the AO representation (site representation)
to the MO one, so that the results can be used for high level post
HF correlated calculations, such as, full CI (FCI), quadruple CI (QCI),
and multi-reference singles-doubles CI (MRSDCI). The other features
of the code include the charge density analysis and the molecular
orbital analysis. Apart from describing the computer program, we also
present several of its applications which include various examples
for $t$-PA, PPP, PPV, acenes, and graphene nanodisks.

The remainder of the paper is organized as follows. In section \ref{sec:theory}
we briefly review the theory associated with the P-P-P model Hamiltonian.
Next, in section \ref{sec:program} we discuss the general structure
of our computer program, and also describe its constituent subroutines.
In section \ref{sec:install} we briefly describe how to install the
program on a given computer system, and to prepare the input files.
Results of various example calculations using our program are presented
and discussed in section \ref{sec:results}. Finally, in section \ref{sec:conclusions},
we present our conclusions, as well as discuss possible future directions.

\section{Theory}

\label{sec:theory}

In this section we briefly review the theory associated with the P-P-P
model Hamiltonian and provide the necessary HF equations within the
basis set approach.

\subsection{Hamiltonian}

The P-P-P Hamiltonian can be written as

\begin{eqnarray}
H_{PPP} & = & \sum_{i,\sigma}\epsilon_{i}c_{i\sigma}^{\dagger}c_{i\sigma}+\sum_{\left\langle ij\right\rangle ,\sigma}t_{ij}(c_{i\sigma}^{\dagger}c_{j\sigma}+c_{j\sigma}^{\dagger}c_{i\sigma})+U\sum_{i}n_{i\uparrow}n_{i\downarrow}\nonumber \\
 &  & +\sum_{i<j}V_{ij}(n_{i}-1)(n_{j}-1)\,,\label{eq:PPP-Hamiltonian}\end{eqnarray}
 where $\epsilon_{i}$ represents the site energy associated with
the $i$th atom, $\left\langle ij\right\rangle $ implies nearest
neighbors, $c_{i\sigma}^{\dagger}$ creates an electron of spin $\sigma$
on the $p_{z}$ orbital of carbon atom $i$ (assuming that the system
is lying in the $xy$-plane), $n_{i\sigma}=c_{i\sigma}^{\dagger}c_{i\sigma}$
is the number of electrons with spin $\sigma$, and $n_{i}=\Sigma_{\sigma}n_{i\sigma}$
is the total number of electrons on atom $i$. The parameters $U$
and $V_{ij}$ are the on-site and long-range Coulomb interactions,
respectively, while $t_{ij}$ is the nearest-neighbor one-electron
hopping matrix element.

The particular forms of the parameters $U$ and $V_{ij}$ determine
the specific variants of the P-P-P model Hamiltonian\cite{sumit-review}.
For $U\neq0$, and $V_{ij}=0$, the model becomes the Hubbard Hamiltonian,
while for $U\neq0$, and $V_{1}\neq0$, but all $V_{ij}=0$, we obtain
the extended Hubbard model. Choosing for the long-range $V_{ij}$
the form

\begin{equation}
V_{ij}=\frac{U}{\left[1+\left(\frac{R_{ij}}{r_{0}}\right)^{2}\right]^{1/2}}\,,\label{eq:V-ohno}\end{equation}
 with $U$ and $r_{0}$ fitted to data gives the Ohno variant\cite{ohno}
of the P-P-P model, whereas taking

\begin{equation}
V_{ij}=\frac{U}{\left[1+\left(\frac{R_{ij}}{r_{0}}\right)\right]}\,,\label{eq:V-mataga}\end{equation}
 gives the Mataga-Nishimoto variant\cite{mataga}. In exponential
variant\cite{exponential}, $V_{ij}$ takes the following form

\begin{equation}
V_{ij}=U\exp\left(-\frac{R_{ij}}{r_{0}}\right)\,,\label{eq:V-exponential}\end{equation}
 In above Eqs.(\ref{eq:V-ohno}, \ref{eq:V-mataga}, \ref{eq:V-exponential}),
$R_{ij}=|\mbox{\textbf{r}}_{i}-\mbox{\textbf{r}}_{j}|$ is the distance
between sites $i$ and $j$ in \AA , while $r_{0}$ is some parameter
in the same units.

In this work, we report calculations based upon the Ohno parametrization
of the P-P-P model mentioned above (\emph{cf}. Eq.\ref{eq:V-ohno}).
Moreover, to account for the inter-chain screening effects we use
a slightly modified form introduced by Chandross and Mazumdar\cite{chandross-mazumdar55},
\begin{equation}
V_{ij}=U/\kappa_{ij}(1+0.6117R_{ij}^{2})^{1/2}\;\mbox{,}\label{eq-ohno6}\end{equation}
 where $\kappa_{ij}$ depicts the dielectric constant of the system
which can simulate the effects of screening and $R_{ij}$ is defined
above. In various calculations performed on phenylene-based conjugated
polymers including PDPAs\cite{shuklamazumdar83,shuklaPRB-62,shukla-synthmet116,shuklaCP-300,shuklaPRB-69,sony-pdpa,shuklaPRB-67,shukla-ppv-synth},
it was noticed that {}``screened parameters'' with $U=8.0$ eV and
$\kappa_{ii}=1.0$, and $\kappa_{ij}=2.0$, otherwise, proposed by
Chandross and Mazumdar\cite{chandross-mazumdar55}, lead to much better
agreement with the experiments, as compared to the {}``standard parameters''
with $U=11.13$ eV and $\kappa_{i,j}=1.0$, proposed originally by
Ohno. In the present code we provide the user with the freedom to
choose these {}``standard'', {}``screened'', or any other user
defined parameters for the Coulomb interactions.

In order to calculate static polarizabilities, one can solve the HF
equations in the presence of an external static electric field. Thus,
we have modified Eq.(\ref{eq:PPP-Hamiltonian}) under the electric
dipole approximation by introducing the corresponding term containing
the uniform electric field $\mathbf{E}$. The overall Hamiltonian
of the system is then given by

\begin{equation}
H_{PPP}^{efield}=H_{PPP}-\mathbf{\boldsymbol{\mu}.E}=H_{PPP}+|e|\mathbf{E.r}\,,\label{eq:PPP-efield-Ham}\end{equation}
 where $H_{PPP}$ is the unperturbed Hamiltonian that describes the
system in the absence of the electric field, $e$ represents the electronic
charge, $\boldsymbol{\mu}=-e\mathbf{r}$, is the dipole operator,
and $\mathbf{r}$ is the positive operator.

\subsection{UHF equations}

Now we describe our formulation for the UHF method. The corresponding
RHF equations can be easily deduced from them. As per the assumptions
of the UHF method, we assume that the $\mu$-th up- and down-spin
orbitals are distinct, and are represented (say) as $\psi_{\mu}^{(\alpha)}$
and $\psi_{\mu}^{(\beta)}$, respectively. We assume that these orbitals
can be written as a linear combination of a finite-basis set \begin{equation}
\psi_{\mu}^{(\alpha)}=\sum_{i}C_{i\mu}^{(\alpha)}\phi_{i},\label{eq:lcao}\end{equation}
 where $\phi_{i}$'s represent the $p_{z}$-orbitals in question,
and the determination of the unknown coefficients $C_{i\mu}^{(\alpha)}$
amounts to the solution of the UHF equations. It is further assumed
that the $\phi_{i}$ orbitals form an orthonormal basis set. In the
equation above, we have only stated the expressions for the up-spin
orbitals, the case of the down-spin orbitals can be easily deduced.
We further assume that the total number of up-/down-spin electrons
is $N_{\alpha}/N_{\beta}$, such that $N_{\alpha}+N_{\beta}=N_{e}$.
Using the conjecture of Eq. (\ref{eq:lcao}) in conjunction with the
Hamiltonian above, one obtains the so-called Pople-Nesbet equations

\begin{equation}
\sum_{j}(F_{ij}^{\alpha}-\varepsilon_{\mu}^{\alpha})C_{j\mu}^{(\alpha)}=0,\label{eq:pople-nesbet}\end{equation}
 where $\epsilon_{\mu}^{\alpha}$ is the UHF eigenvalue of the $\mu$-th
up-spin orbital, $F_{ij}^{\alpha}$ is the Fock matrix for the up-spin
electrons defined by the equations\begin{equation}
F_{ij}^{\alpha}=t_{ij}-P_{ij}^{\alpha}V_{ij},\,\,\,(i\neq j)\label{eq:Fock_ij}\end{equation}
 \begin{equation}
F_{ii}^{\alpha}=\epsilon_{i}-\sum_{j\neq i}V_{ij}+\sum_{j}P_{jj}V_{ij}-P_{ii}^{\alpha}V_{ii},\,\,\,(i=j)\label{eq:fock}\end{equation}
 where $\epsilon_{i}$, $t_{ij}$ and $V_{ij}$ are defined above
(\emph{cf.} Eq. \ref{eq:PPP-Hamiltonian}), and $P_{ij}^{\alpha}$
and $P_{ij}$, are the up-spin and total density matrix elements,
respectively, defined as \begin{equation}
P_{ij}^{\alpha}=\sum_{\mu=1}^{N_{\alpha}}C_{i\mu}^{(\alpha)*}C_{j\mu}^{(\alpha)},\label{eq:den-up}\end{equation}
 and\begin{equation}
P_{ij}=P_{ij}^{\alpha}+P_{ij}^{\beta}.\label{eq:den-tot}\end{equation}
 Once the Fock matrix is constructed, one solves the eigenvalue problem
for the up-spin Fock matrix (\emph{cf}. Eq. \ref{eq:pople-nesbet})
as well as the down-spin Fock matrix, using the iterative diagonalization
technique, to achieve self-consistency. From the equations given above,
it is easy to deduce the expressions for $F_{ij}^{\beta}$, as well
as the Fock matrix elements for the RHF case.

\section{Description of the Program}

\label{sec:program}

Our computer code consists of the main program, and various subroutines,
all of which have been written in Fortran 90 language. Additionally,
the program must link to the LAPACK/BLAS library, whose diagonalization
routines are used by our program.

Before we describe various subroutines in detail, it is fruitful to
discuss the scaling of the memory requirements, and the cpu time (calculation
time), with the system size. If $N$ is the number of sites ($\pi$
orbitals) in the system, then the memory requirements will roughly
scale as $N^{2}$, needed mainly for the storage of the Fock matrix,
and its eigenvectors. The cputime, on the other hand, will be dominated
by the self-consistency iterations (the so-called {}``self-consistent
field'' (SCF) process), which consists of construction of the Fock
matrix, and its diagonalization. The cputime needed for the construction
of the Fock matrix scales as $N^{2}$, while the diagonalization time
exhibits $N^{3}$ scaling. Thus, we conclude that the dominant scaling
of the cputime is $N^{3}$, with respect to the system size. For the
open-shell systems, for which the UHF procedure is needed, the memory
and cpu time requirements will roughly be twice as compared to a closed
shell system of similar size because the UHF procedure involves the
construction and diagonalization of two Fock matrices, corresponding
to the {}``up'', and the {}``down'' spins.

Several molecules in nature are highly symmetric, and, therefore,
in general, it is fruitful to implement the point-group symmetries
in electronic-structure codes like the present one. However, our code
does not make use of point-group symmetries mainly because the computation
time associated even for large systems is really insignificant. Therefore,
the programming effort required to implement various irreducible representations
will simply outweigh the rewards in terms of reduced cpu time. Moreover,
the irreducible representations associated with various orbitals and
many-particle states can be easily determined by examining the transition
dipole moments among them, in conjunction with the dipole-selection
rules of various point groups. Next, we briefly describe the main
program, and each of the subroutines of our code.

\subsection{Main program PPP}

This is the main program of our package which reads input data such
as which Hamiltonian to use, its parametrization, charge on the system,
total number of atoms in the unit cell, and their coordinates. The
program also reads in the options to perform various types of calculations
such as RHF, UHF, SCI, optics, etc. The option is also provided to
perform HF calculations in the presence of an external electric field,
which is specified by its Cartesian components in the units of V/\AA.
The program can delete the atoms and calculates the entities like
total number of sites, electrons and occupied orbitals. It dynamically
allocates various arrays, and then calls other subroutines to accomplish
the remainder of the calculations. Because of the dynamical array
allocation, the user need not worry about various array sizes, as
the program will automatically terminate when it exhausts all the
available memory on the computer.

\subsection{Subroutine R\_ATOM}

This subroutine reads the coordinates of the atoms of a unit cell
with respect to its user specified origin. It can also generate coordinates
of some important structural units such as a bond, phenyl group (benzene),
and even fullerene (C$_{60}$) to facilitate an easy realization of
the molecular system under consideration.

\subsection{Subroutine BENPERP}

This routine generates the coordinates of six carbon atoms forming
the benzene backbone. The orientation is perpendicular to the conventional
orientation. Note that the origin of the benzene skeleton is considered
to be the center of the hexagonal ring. Rotations, if desired, are
performed keeping the origin fixed. Positive angles are considered
to be anti-clockwise, while the negative angles are treated clockwise.
Finally, the center is translated to the location specified by the
user. The plane in which benzene locates is taken as a card (XY or
YZ or ZX) in the input file.

\subsection{Subroutine BENZEN}

Similar to subroutine BENPERP, this routine also generates the coordinates
of six carbon atoms forming the benzene backbone. This routine differ
from the previous one in a way that it creates benzene ring in the
orientation parallel to the conventional orientation. One can also
rotate and translate the ring, and can also specify the plane in which
benzene ring lies.

\subsection{Subroutine BOND}

This routine generates a bond (2 carbon atoms). Center of the bond
is taken to be the origin and initially the bond is assumed to be
along the x-axis. Rotations, if desired, are performed keeping the
origin fixed. Positive angles are considered for anti-clockwise rotation,
while the negative angles are treated for clockwise rotation. Finally,
the center is translated to the location specified by the user.

\subsection{Subroutine STLINE}

This subroutine also generates a bond, but with the first atom of
the bond at the origin. Here, initially the bond is assumed to be
along the x-axis. Operations like rotation and translation can also
be performed.

\subsection{Subroutine C60\_GEN}

This routine, which is originally written by Dharamvir \textit{et
al}.\cite{c60-coord}, generates the position coordinates of all 60
carbon atoms of the fullerene structure. Inputs are two bond lengths,
\textit{i.e.}, single and double bond lengths. This does not have
the {}``standard'' orientation, but one with pentagons at top and
bottom.

\subsection{Subroutine R\_SITE}

If the molecule under consideration consists of translationally invariant
units, this routine generates the coordinates of all the $\pi$-electron
sites in the system from the coordinates of origins of the reference
unit cell and the translational vector of the lattice. At present
this subroutine is restricted to periodicity only in one dimension,
thus, making it useful for polymers and their oligomers.

\subsection{Subroutine DELATM}

This subroutine deletes the users specified atoms from the list containing
all the atomic coordinates.

\subsection{Subroutine SYMSITE}

This routine symmetries the coordinates of the atoms to place the
origin of the molecule at the center of mass of the molecule.

\subsection{Subroutine PRINTR}

This routine truncates the position coordinates to seven places to
right of the decimal and then prints out the coordinates of all the
sites into the output file. In addition, it also creates a file named
as 'atomic\_coord.xsf', which can be used as input file in XCrySDen\cite{xcrysden}
program, meant for visualizing the molecular structure of the system
under consideration.

\subsection{Subroutine MATEL}

This is the master routine meant for generating the one-electron ($t_{ij}$)
and two-electron ($V_{ij}$) matrix elements, with or without the
electric field. This is done based upon the data specified by the
user in the input file such as the Hamiltonian under consideration,
Coulomb parameters to be used (if any), hopping matrix elements connecting
various sites, \textit{etc}.

\subsection{Subroutine Tij\_READB}

This routine is meant for reading the hopping related input for the
band structure calculations using the TB model (Hückel model). It
reads the unique intracell, as well as intercell, hopping matrix elements,
and their connectivities from the input file. Note that an one dimensional
tight-binding system with only nearest-neighbor coupling is assumed.
Thus, a given cell can only connect to at the most to two neighbors
(one to the left and the other to the right). For anything else the
code will have to be modified. Also one has to specify the unit cell
to which it is connected.

\subsection{Subroutine READ\_EI}

If there any site energies are, this routine reads them. In case of
more than one unit cell, the routine also generates rest of the site
energies using translational invariance.

\subsection{Subroutine Tij\_READ}

This subroutine reads the unique hopping matrix elements and their
connectivities from the input file. In case the system consists of
more than one unit cell, the routine also generates rest of the hoppings
from translational invariance.

\subsection{Subroutine Tij\_GEN}

This routine is used in order to generate hopping automatically from
the information of bond distance and the corresponding hopping matrix
elements. The subroutine considers all pairs of sites in the molecule
and if the distance between them is equal to the user provided distance,
it assigns to them the user provided hopping matrix elements. This
automates the generation of hopping matrix elements to a great extent,
thus, simplifying the input data. This routine is invoked by the card
'HOPGEN' in the input file.

\subsection{Subroutine Vij\_CAL}

This subroutine computes the long-range Coulomb matrix elements $V_{ij}$
(\textit{cf}. Eqs. \ref{eq:V-ohno}, \ref{eq:V-mataga}, \& \ref{eq:V-exponential})
for the P-P-P model with three possible parametrization namely Ohno,
Mataga-Nishimoto, or exponential, depending upon the input provided.
In case the choice of Hamiltonian is the Hubbard or the extended Hubbard
model, the relevant Coulomb parameters $U$ and/or $V$, are provided
in the input itself. In case of RHF calculations, the routine can
also calculate the contribution corresponding to the correlation functions,
if such a calculation has been requested in the input file. For any
choice of a correlated Hamiltonian, corresponding interaction matrix
elements are stored in an array.

\subsection{Subroutine BANDS}

This is the subroutine which performs the band structure calculations
using the Hückel model. In order to get the band structure, first
Fourier transformation of the hopping matrix elements is performed,
and later the matrix is diagonalized using the subroutine ZHPEV from
the LAPACK/BLAS library to obtain the eigenvalues and eigenvectors.
The eigenvalues of corresponding bands are written in an ASCII file
'bands.dat' which can be used for plotting the bands using standard
programs such as xmgrace\cite{xmgrace} or gnuplot\cite{gnuplot}.
The eigenvectors are written in the binary format, in the file named
'bloch\_orbitals.dat'.

\subsection{Subroutine FOUTRA\_TB}

This subroutine computes the Fourier transform of one dimensional
nearest-neighbor-interacting tight-binding Hamiltonian. It is called
by subroutine BANDS, discussed above.

\subsection{Subroutine SCF\_RHF }

This subroutine solves the RHF equations for the system under consideration
in a self-consistent manner, using the iterative diagonalization procedure
and returns the canonical SCF orbitals, their eigenvalues, and the
total energies. The arrays which are needed during the calculations
are allocated before the calculations begins, and are deallocated
upon completion. Before the first iteration, Hückel model Hamiltonian
is diagonalized to obtain a set of starting orbitals. Subsequently,
the Fock matrix corresponding to those orbitals is constructed and
diagonalized. The process is repeated until the self-consistency is
achieved. During the self-consistency iterations, subroutine DSPEVX
from the LAPACK/BLAS library is used to obtain the occupied eigenvalues
and eigenvectors. Obtaining only the occupied eigenpairs, as against
the entire spectrum, leads to considerable savings of CPU time for
large systems. However, if the entire spectrum of eigenvalues and
eigenvectors is needed, say, to perform optical absorption calculations
or preparing the data for subsequent correlation calculations, the
Fock matrix is diagonalized using the routine DSPEV from the LAPACK/BLAS
library, upon completion of the self-consistency iterations. Because
the entire spectrum is obtained only once the self-consistency has
been achieved, it does not strain the computational resources too
much.

\subsection{Subroutine SCF\_UHF }

This subroutine is exactly the same in its logic and structure as
the previously described SCF\_RHF, except that the task of this routine
is to solve the UHF equation for the system under consideration. Different
Fock matrices for the up- and the down-spin are constructed and diagonalized
in each iteration, until the self-consistency is achieved. Similar
to the case of routine SCF\_RHF, during the iterations only the occupied
eigenvalues and eigenvectors are computed using the routine DSPEVX.
The iterations are stopped once the total UHF energy of the system
converges to within a user defined threshold.

\subsection{Subroutine DENSITY}

This subroutine constructs the density matrix for the closed-shell
systems, assuming orbitals to be doubly occupied.

\subsection{Subroutine DENSITY\_UHF}

This subroutine constructs density matrices needed for the UHF calculations.
It generates different density matrices for the orbitals with up ($\alpha$)
and down ($\beta$) spins, and in the end adds them to compute the
total density matrix.

\subsection{Subroutine FOCKMAT}

This subroutine constructs the Fock matrix for the closed-shell system
and calculates the total SCF energy from the density matrix and the
one- and two-electron integrals.

\subsection{Subroutine FOCKMAT\_UHF}

This subroutine is analogous to the routine FOCKMAT, the only difference
is that it constructs the Fock matrix for the UHF calculations by
computing separate Fock matrices for up- and down-spins. Thereafter,
it calculates the total SCF energy from the density matrices and the
one- and two-electron integrals.

\subsection{Subroutine ORBDEN}

This routine computes the charge density of a given set of orbitals
on user specified sites. This helps in understanding as to how the
charge is distributed in an extended system and helps us to visualize
that whether a given orbital is localized or delocalized. This analysis
can be performed both on RHF and UHF orbitals.

\subsection{Subroutine WRITORB}

The purpose of this subroutine is to write the converged orbitals
to the orbital file in the ASCII format. In case of RHF calculations,
orbitals will be written in 'ORB001.DAT' file, while in case of UHF
calculations, orbitals with up- and down-spins will be written in
the files 'ORBALPHA.DAT' and 'ORBBETA.DAT', respectively.

\subsection{Subroutine SCI}

This subroutine is the master routine for performing the single configuration
interactions (SCI) calculations, using the Hartree Fock occupied and
virtual MOs. SCI is a simple approach which accounts for the electron-correlation
effects by including configurations which are singly-excited with
respect to the HF reference state, in the CI expansion\cite{salem}.
Because of the Brillouin theorem, which forbids the mixing of the
HF state with the singly-excited configurations, the ground state
remains unchanged in a SCI calculation\cite{salem}. However, excited
configurations do mix with each other leading to a reasonably good
description of the excited states in this approach\cite{salem}. In
the present version of code, SCI method is implemented only for the
case of closed-shell systems, for which RHF calculations are performed
prior to the SCI ones. To obtain the CI eigenvalues, SCI matrix is
diagonalized using the routine DSPEV from the LAPACK/BLAS library.

\subsection{Subroutine HAM\_SCI}

This routine constructs the single-CI Hamiltonian matrix for the singlet
subspace. A closed-shell reference state is assumed. The Hamiltonian
matrix elements between different sets of configurations are computed
using the standard Slater-Condon rules\cite{salem}.

\subsection{Subroutine OPTICS}

This is the master routine meant for computing the optical absorption
spectrum, which can be obtained at the single particle level (TB or
HF) as well as at the SCI level. The range of frequencies over which
the spectrum is to be computed, along with the line width, are read
from the input file. The calculated spectrum is written in an output
file called 'spec001.dat', which can be plotted using xmgrace\cite{xmgrace}
or gnuplot\cite{gnuplot}. At present this routine is restricted to
the closed-shell systems.

\subsection{Subroutine SPCTRM\_A}

This routine computes the absorption spectrum from the ground state
under electric-dipole approximation, assuming a Lorentzian line shape
and a constant line width for all the levels. The computed spectrum
is written in an ASCII file 'spec001.dat', which can be readily used
for plotting using programs such as xmgrace\cite{xmgrace} or gnuplot\cite{gnuplot}.

\subsection{Subroutine SPCTRM\_1EX}

This is an important subroutine which calculates the linear optical
absorption of the system from a singly-excited state (as compared
to the ground state) for one electron TB or HF calculation. The formalism
as well as the approach of computation is same as that in routine
SPCTRM\_A, and the output of this routine is also written in the ASCII
file 'spec001.dat'. Excited state absorption spectra are useful in
interpreting photo-induced absorption experimental data.

\subsection{Subroutine DIPMAT}

If optical absorption calculations are to be performed within the
SCI approach, one needs electric dipole matrix elements between various
singly excited configurations. The task of this routine is to compute
these matrix elements so that they can be used for calculating the
optical absorption spectrum in the routine SPCTRM\_A.

\subsection{Subroutine NLO}

This subroutine generates the input file to be used for the nonlinear
optical (NLO) susceptibility calculations at the single particle level.
In order to reduce the size of dipole matrix elements, one can delete
the outermost occupied and virtual orbitals choosing the card 'ORBDEL'.
The program deletes the orbitals in the given range using a subroutine
named 'delo\_nlo', which will be described next. Output data of NLO
calculations is written in file 'NLO001.DAT', which includes total
number of orbitals, number of occupied orbitals, energies of the occupied,
as well as virtual states, and the dipole moments. NLO option is also
restricted to the RHF calculations only. This data can be used to
compute NLO susceptibilities such as two-photon absorption (TPA),
third harmonic generation (THG), \textit{etc.}, using the sum-over-states
(SOS) formalism of Orr and Ward\cite{OrrWard}.

\subsection{Subroutine DELO\_NLO}

The task of this subroutine is to delete occupied and virtual orbitals
from the orbital array for the NLO calculations. This routine is called
by subroutine NLO, if needed.

\subsection{Subroutine DIPCAL}

This routine computes the dipole moment matrix between the single
particle orbitals under the Hückel approximation, according to which,
among the AOs only diagonal matrix elements are nonzero, and their
values are nothing but their Cartesian locations. This routine simply
transforms the dipole matrix elements from the AO representation to
the MO one. This routine is called from the routine NLO and its aim
is to supply dipole integrals needed for subsequent SOS calculations
of NLO susceptibilities mentioned above.

\subsection{Subroutine CI\_DRV}

This is the driver routine which controls the preparations of various
data files needed for the post-HF correlated calculations, such as
FCI, QCI, MRSDCI, \textit{etc}. The tasks of this routine include: 
\begin{enumerate}
\item Reading the orbitals 
\item Freezing and/or deleting orbitals if specified by the user in the
input file 
\item Creating the one-electron effective core potential (ECP) corresponding
to the frozen set of orbitals, if needed 
\item Transforming the one- and two-electron integrals from the site (AO)
representation to the basis spanned by these orbitals (MOs). They
are written to different files so that they can be read and used by
a separate post-HF correlation program. 
\end{enumerate}

\subsection{Subroutine ECP}

This program calculates the effective one-electron frozen core potential
matrix elements, and stores them in the array. Eventually, these matrix
elements are added to the other one-electron part of the Hamiltonian
before being written out in the CI\_DRV output files. The energy contribution
of the frozen orbitals is stored in the variable \emph{ecore}.

\subsection{Subroutine TWOIND}

This routine performs the two-index transformation on one-electron
matrix elements, so as to transform them from the AO to the MO representation
for the use in subsequent correlation calculations. It is called by
the master routine CI\_DRV.

\subsection{Subroutine WRITE\_1}

This routine writes all the nonzero one-electron Hamiltonian matrix
elements expressed in the MO representation in a binary file called
'ONEINT001.DAT'. This routine is also called by the master routine
CI\_DRV, and the output is meant to be used in the post-HF correlation
calculations.

\subsection{Subroutine FOURIND}

This routine performs the four-index transformation on the two-electron
integrals so as to transform them from the AO to the MO represention.
This routine is also called by the master routine CI\_DRV.

\subsection{Subroutine WRITE\_2}

This routine writes all the nonzero two-electron Hamiltonian matrix
elements expressed in the MO representation in a binary file called
'TWOINT001.DAT'. This routine is also called by the master routine
CI\_DRV, and the output is meant to be used in the post-HF correlation
calculations.

\subsection{Subroutine DIPINT}

This routine transforms the dipole operator from the site (AO) representation
into the representation of the orbitals (MOs). The routine is called
from CI\_DRV and it provides dipole matrix elements to be used for
optical properties calculations in conjunction with post-HF correlated
calculations. The algorithm used in this routine is the one described
in the routine DIPCAL.

\subsection{Subroutine DIPOUT}

The purpose of this routine is to write all the nonzero dipole matrix
elements in the binary file 'DIPINT001.DAT' to be used in conjunction
with the post-HF correlated calculations. This routine is also called
from CI\_DRV.

\section{Installation, input files, output files}

\label{sec:install}

We believe that the installation and execution of the program, as
well as preparation of suitable input files is fairly straightforward.
Therefore, we will not discuss these topics in detail here. Instead,
we refer the reader to the README file for details related to the
installation and execution of the program. Additionally, the file
'\texttt{manual.pdf}' explains how to prepare a sample input file.
Several sample input and output files corresponding to various example
runs are also provided with the package.

\section{Results and Discussions}

\label{sec:results}

In this section, we present and discuss the numerical applications
of our code. First we present the results on the convergence of total
energy with respect to the oligomer's length for various conjugated
polymers such as $t$-PA, PPP, PPV, and polyacene. We also provide
our results for the UHF calculation on graphene nanodisks. Next, we
present our results for the tight binding band structure calculations
for PPP. Further, we discuss and compare the linear absorption optical
spectrum of PPP computed by different approaches such as TB, HF, and
SCI. As mentioned in the above sections that the program can perform
calculations in the presence of finite electric field as well, which
makes it capable of calculating electro-absorption spectrum. For centro-symmetric
systems, the electro-absorption spectrum allows one to explore both
the even and the odd parity excited states in a linear optical absorption
process. Results and analysis of the electro-absorption spectrum of
an oligomer of PPP at the SCI level are also presented.

\subsection{Total Energy}

\label{subsec:total energy}

Our code can be used to study both the ground and the excited state
properties of oligomers of various polymers because they are nothing
but finite molecules, ranging in size from small to large. However,
in this section we demonstrate that our code can also be used to obtain
the ground state energy/cell, in the bulk limit, for one-dimensional
periodic systems such as polymers.

The energy per unit cell of a one-dimensional periodic system can
be obtained using the formula

\begin{center}
\begin{equation}
E_{cell}=\lim_{n\rightarrow\infty}\Delta E_{n}=\lim_{n\rightarrow\infty}(E_{n+1}-E_{n}),\label{eq:ecell}\end{equation}
 
\par\end{center}

where $E_{n+1}/E_{n}$ represent the total energies of oligomers containing
$n+1/n$ unit cells. Thus, using this formula, for sufficiently large
value of $n$, one can obtain the energy/cell of a polymer in the
bulk limit, from oligomer based calculations. In what follows we show
that value of $E_{cell}$ converges quite rapidly with respect to
$n$.

\begin{figure}
\begin{centering}
\includegraphics[width=12cm]{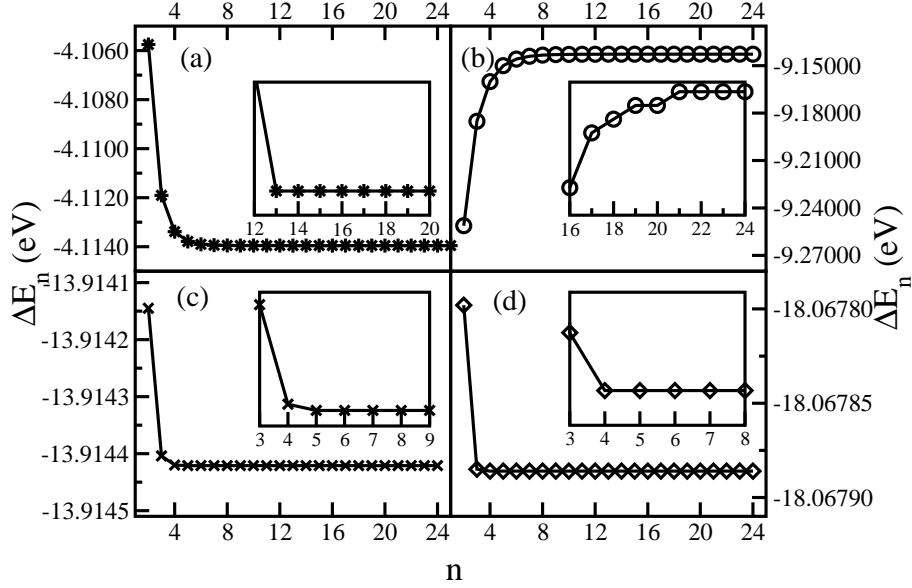}
\par\end{centering}

\caption{Convergence in $\triangle E_{n}$ with respect to the oligomer length
($n$) of various systems: (a) $t$-PA, (b) polyacene, (c) PPP, and
(d) PPV. Inset of each graph shows the zoomed plot, in the energy
window of $10^{-7}$ -- $10^{-8}$~eV.\label{fig:Convergence} }

\end{figure}

First we examine the convergence of $E_{cell}$ obtained using Eq.
(\ref{eq:ecell}) with respect to the number of unit cells $n$, using
the RHF approach. The energy convergence threshold for the SCF calculations
is set up to eighth decimal place. In Fig. \ref{fig:Convergence}
we plot $\Delta E_{n}$ as a function of $n$, for various polymers
such as $t$-PA, acene, PPP, and PPV. From plots (c) and (d) in Fig.
\ref{fig:Convergence}, it is evident that energy for PPP and PPV
converges rapidly and convergence is achieved for short oligomer length
(6--7 units), while in $t$-PA and acene (Fig. \ref{fig:Convergence}(a)
and (b), respectively), the convergence is achieved for larger value
of $n$.

\begin{figure}
\begin{centering}
\includegraphics[width=8cm]{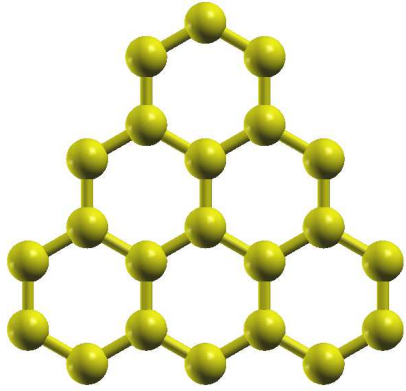}
\par\end{centering}

\caption{Schematic diagram of the trigonal zigzag graphene nanodisk with six
benzene rings (22 sites) considered in this work.\label{fig:benzo6} }

\end{figure}

The above mentioned calculations can also be performed using the UHF
approach. However, it is more meaningful to use this approach for
systems with open-shell configurations, such as trigonal zigzag graphene-nanodisks.
Thus, UHF calculations can be used to ascertain whether or not a high-spin
state is the ground state. For example, we found that the ground state
of a trigonal zigzag graphene-nanodisk with six benzene rings (\emph{cf}.
Fig. \ref{fig:benzo6}) is a triplet one (see Table \ref{tab:UHF}).

\begin{table}
\caption{Results of UHF total energy calculations on a trigonal zigzag graphene
nanodisk with six benzene rings (\emph{cf.} Fig. \ref{fig:benzo6})
for various combinations of majority ($n_{\alpha}$) and minority
($n_{\beta}$) spins. \label{tab:UHF} Our results indicate that the
triplet state being significantly lower in the total energy than the
singlet state, is the true ground state.}

\centering{}\begin{tabular}{|c|c|c|}
\hline 
$n_{\alpha}$  & $n_{\beta}$  & $E$ (eV)\tabularnewline
\hline
\hline 
11  & 11  & -49.840415300708\tabularnewline
\hline 
12  & 10  & -51.858572334959\tabularnewline
\hline 
13  & 9  & -48.083423107736\tabularnewline
\hline 
14  & 8  & -43.974251795664\tabularnewline
\hline
\end{tabular}
\end{table}

\subsection{Band Structure}

\label{subsec:BS}

\begin{figure}
\vspace{2cm}

\begin{centering}
\includegraphics[width=12cm]{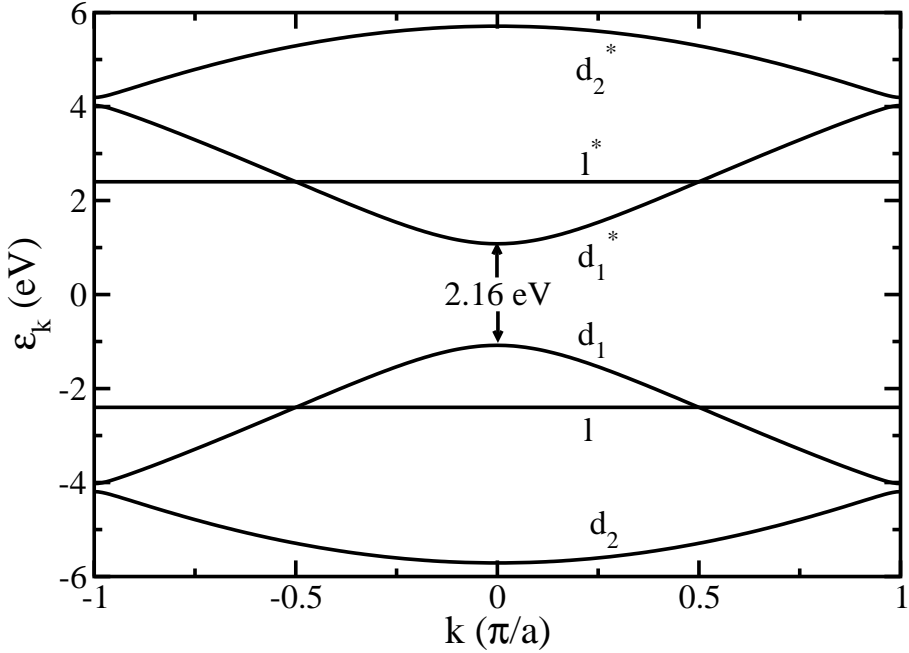}
\par\end{centering}

\caption{Band structure of PPP, computed using Hückel model Hamiltonian. Dispersion
less flat bands denote the localized levels ($l/l^{*}$), while the
other bands represent the delocalized ones. \label{fig:Band-Structure}}

\end{figure}

Our code can be used to perform band structure calculations using
the nearest-neighbor TB model. In Fig. \ref{fig:Band-Structure},
we present band structure of PPP. It possesses six bands out of which
two are localized (labeled as '$l$', '$l^{*}$'), while four are
delocalized (labeled as '$d_{1}/d_{2}$', '$d_{1}^{*}/d_{2}^{*}$').
As is obvious from the Fig. \ref{fig:Band-Structure}, the band structure
has a direct band gap, $E_{g}=2.16$ eV at the $\Gamma$-point. Our
calculated band structure is fully consistent with that reported by
Heeger and co-workers\cite{Heeger-PPP}.

\subsection{Linear optical absorption spectrum}

\label{subsec:OA}

In order to understand the optical properties of conjugated polymers,
it is essential to study their low-lying excited states. The linear
optical absorption spectrum, which is all about one-photon absorption,
provides an efficient medium to explore the odd-parity low-lying states
in centro-symmetric systems. In an independent particle model it can
be explained in terms of excitations from ground state, by promoting
one electron from one of the occupied MO (valence band states) to
one of the unoccupied MO (conduction band states). While, in a correlated
electron approach, such as the P-P-P model, this one-electron picture
does not hold, and the excited states are linear combination of several
configurations obtained by performing CI calculations.

Our code can be used to compute linear optical absorption spectrum
of $\pi$-conjugated materials using the Hückel model (TB approach),
as well as P-P-P model Hamiltonians employing both the HF and the
SCI approaches.

\begin{figure}
\begin{centering}
\includegraphics[width=12cm]{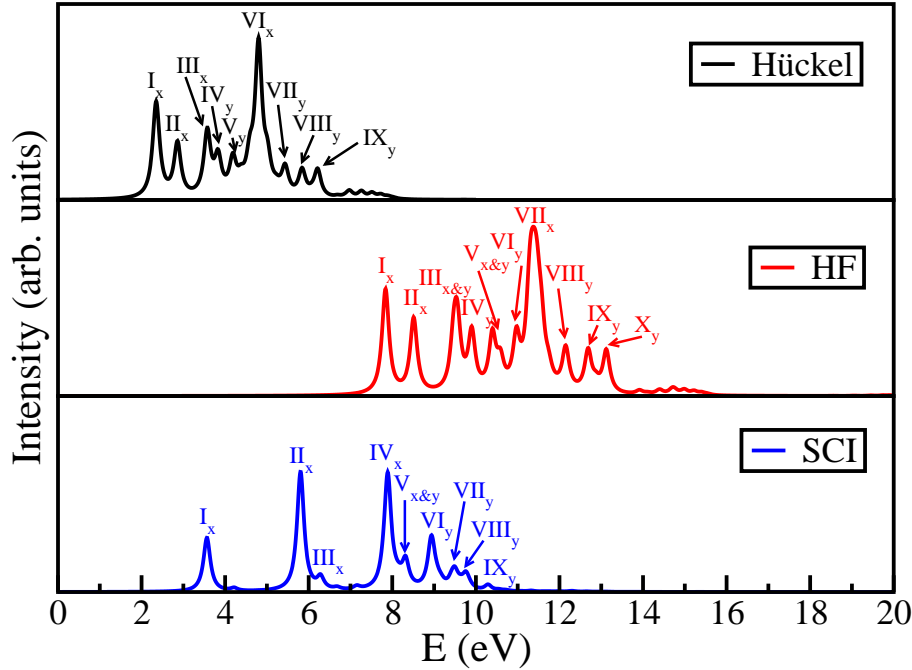}
\par\end{centering}

\caption{Linear optical absorption spectrum of planar PPP-8, computed using
the Hückel as well as P-P-P models, both at the HF and the SCI levels.
Various peaks and their polarizations are labeled. The spectra are
plotted using the line width of 0.1 eV. \label{fig:optical-absorption}}

\end{figure}

In Fig. \ref{fig:optical-absorption} we present the linear optical
absorption spectrum of planar PPP with eight repeat units (PPP-8)
using TB, HF, and SCI approaches. The spectrum was calculated using
the carbon-carbon bond length (within a phenyl ring) to be 1.4 \AA,
while the inter-ring single bond length was taken to be 1.54 \AA.
The standard value of -2.4 eV was used for the intra-benzene hopping,
while for the inter-ring single bond we used the value -2.23 eV. The
HF and the SCI calculations were performed using the standard parameters
in the P-P-P Hamiltonian. All the three graphs depicted in Fig. \ref{fig:optical-absorption}
are qualitatively quite similar, with the first two features, and
the most intense feature, corresponding to the $x$-polarized transition.
The spectrum drawn at the HF level using P-P-P model is blue shifted
as compared to the Hückel model spectrum, which is due to the well
known tendency of the HF approach to over estimate the band gaps.
However, inclusion of correlation effects at the SCI level brings
the spectrum back to lower energy range. The location of the lowest
peak corresponds to the optical gap, which is found to be around 3.5
eV, in excellent agreement with the reported first-principles calculations\cite{cad-PPP},
and also in good agreement with the experimental value ($\approx$3
eV) for the infinite polymer\cite{athouel-exp}. A higher level CI
calculations like FCI, QCI, and MRSDCI can also be performed using,
\textit{e.g.}, the MELDF code\cite{meld} beyond HF. The input for
these higher CI calculations can be prepared using CIPREP option in
the input file.

On analyzing the wave functions of the excited states contributing
to the spectra, we find that first peak in TB and HF spectra is due
to transition from the highest occupied molecular orbital (HOMO) to
the lowest unoccupied molecular orbital (LUMO), \textit{i.e.}, $H\rightarrow L$.
In the spectrum obtained using SCI approach, for peak I (see Fig.
\ref{fig:optical-absorption}), the major contribution comes from
$H\rightarrow L$ configuration, while the contributions due to $H-1\rightarrow L+1$,
$H-2\rightarrow L+2$, $H-3\rightarrow L+3$, and $H-12\rightarrow L+12$
are also significant.

\begin{figure}
\vspace{1cm}

\begin{centering}
\includegraphics[width=12cm]{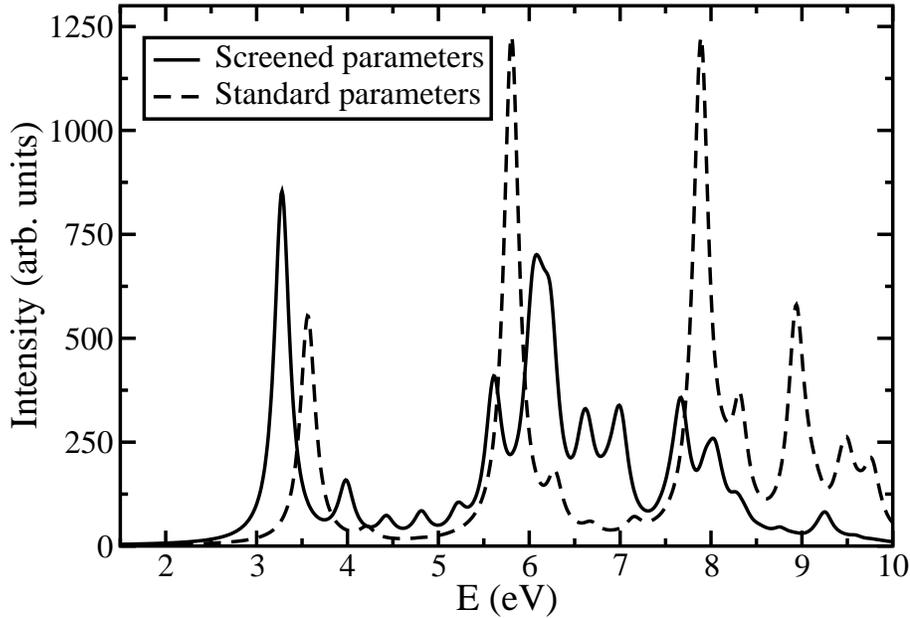}
\par\end{centering}

\caption{Linear optical absorption spectrum of PPP-8 computed using the standard
(dashed line) as well as the screened (solid line) set of Coulomb
parameters in the P-P-P model, using the SCI approach. The spectra
are plotted using the line width of 0.1 eV. \label{fig:Linear-optical-spectrum}}

\end{figure}

Our code is also capable of examining the role of Coulomb parameters
used in the P\textendash{}P\textendash{}P Hamiltonian by not only
considering standard and screened set of parameters defined earlier
(\emph{cf}. Sec. \ref{sec:theory}), but also any user defined set
of Coulomb parameters, which may be material specific. Fig. \ref{fig:Linear-optical-spectrum}
depicts the linear absorption spectra computed using standard parameters
(dashed line) and screened parameters (solid line), for planar geometry
of PPP-8 at SCI level. Both the spectra are qualitatively similar,
while quantitatively, spectrum computed using screened parameters
is red shifted as compared to the standard parameter spectrum. We
note that the value 3.27 eV of the optical gap obtained using the
screened parameters is in better agreement with the experimental value
($\approx3$eV) for the infinite system\cite{athouel-exp}.

\subsection{Electro-absorption spectrum}

\label{subsec:EA}

As mentioned earlier our code can also solve the HF equations in the
presence of a finite electric field. This allows us, in principle,
to use it for computing quantities such as the static dielectric polarizabilities
and the electro-absorption (EA) spectra of various conjugated systems.
For example, the first-order polarizability is determined as the second
derivative of the total energy per unit cell ($E_{total}$) with respect
to the strength of the external electric field $\mathbf{E}$. The
second derivative of $E_{total}$ can be calculated employing finite
difference formulas, using $E_{total}$ with, and without, the electric
field. However, in this work we are concentrating on another possible
application of the finite-field approach, namely the electro-absorption
spectrum.

The electro-absorption spectrum is an important experimental probe
for various materials which, through linear optical absorption allows,
one to probe both even, as well as odd parity states for centro-symmetric
systems. The noteworthy point is that without electro-absorption normally
one will have to use both linear absorption as well as nonlinear absorption
to investigate states of both parities. Using our program one can
easily make theoretical prediction of this quantity using its definition:
$\Delta\alpha(\omega,\mathbf{E})=\alpha(\omega,\mathbf{E})-\alpha(\omega,0)$,
where $\alpha(\omega,\mathbf{E})$ is the optical absorption spectrum
in the presence of the external electric field $\mathbf{E}$. In order
to compute it, the HF equations are solved in the presence of the
field $\mathbf{E}$ (\emph{cf.} Eq. \ref{eq:PPP-efield-Ham}) leading
to a modified set of orbitals and their eigenvalues, as compared to
the zero-field ($\mathbf{E}=0$) case. This orbital set can subsequently
be used to compute: (a) optical absorption at the HF level, or (b)
optical absorption at any level of correlation. Then by using the
above mentioned definition of $\Delta\alpha(\omega,\mathbf{E})$ one
can compute the corresponding EA spectrum. In this work we present
the results of the EA spectrum of PPP-8, computed at the SCI level
using our program. The EA spectra are usually interpreted in terms
of Stark shift analysis and it roughly follows the first derivative
of linear absorption, calculated in the presence of the electric field,
with respect to the photon energy.

\begin{figure*}
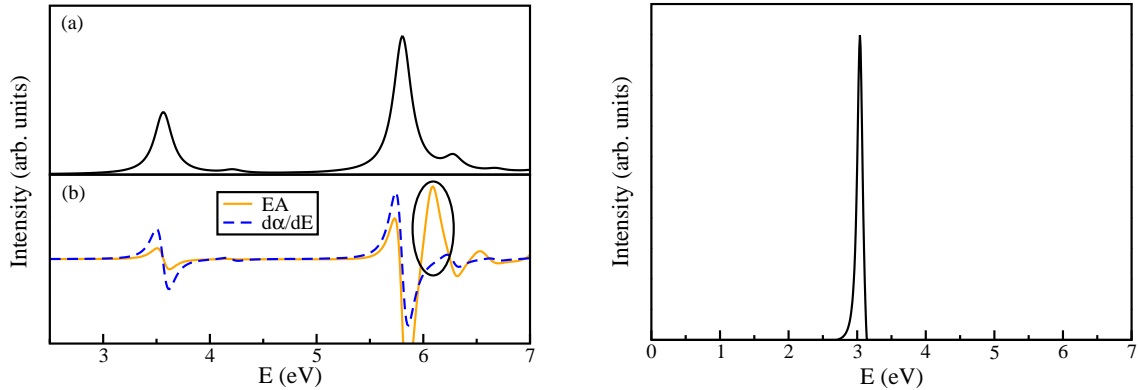

\includegraphics[width=7cm]{fig6a}\hspace{1cm}\includegraphics[width=7cm]{fig6b}

\caption{Left panel: (a) The linear absorption spectrum planar molecule of
PPP-8, computed in the presence of finite electric field, using standard
parameters in the P-P-P Hamiltonian with the SCI approach. (b) Electro-absorption
spectrum (solid line) and the first derivative of linear absorption
with respect to the photon energy $\mbox{E}$ (dashed line) for the
same system. Right panel: The two photon absorption (TPA) spectrum
of PPP-8, computed at SCI level, using the MELDF program\cite{meld}.
In all the figures above, E represents the photon energy. The $mA_{g}$
state in the TPA spectrum corresponds to the encircled feature in
the EA spectrum, as is also clear by the peak positions. \label{fig:EA}}

\end{figure*}

Fig. \ref{fig:EA} (a) shows the linear absorption spectrum computed
in the presence of finite electric field, using the standard parameters
in the P-P-P Hamiltonian with the SCI approach. As expected, the most
intense features in the EA spectrum appears when the field is applied
along the conjugation direction which happens to be the $x$-axis
in the present case. In these calculations the electric field strength
was taken to be $10^{-3}$ V/\AA, which is of the same order of magnitude
used in various experiments\cite{Vardeny-EA}. Fig. \ref{fig:EA}
(b) depicts the EA spectrum together with the first derivative of
linear absorption with respect to the photon energy $\mbox{E}$. We
note that the EA spectrum resembles the first derivative of linear
absorption spectrum, roughly. The first absorption peak in linear
absorption spectrum occurs at 3.56 eV, following the other features
at 4.21 eV, 5.8 eV, and 6.27 eV {[}see Fig. \ref{fig:EA} (a){]}.
In EA spectrum, the first peak appears at 3.51 eV, which is similar
to $\mbox{d\ensuremath{\alpha}/dE}$ spectrum {[}dashed curve in Fig.
\ref{fig:EA} (b){]}. We therefore assign this feature to a Stark
shift of the lowest odd parity ($1B_{u}$) exciton. There is also
a feature at 6.09 eV {[}circled in Fig. \ref{fig:EA} (b){]}, which
cannot be found by a derivative analysis of the absorption spectrum,
and must therefore be due to an even parity ($mA_{g}$) exciton. To
verify this, we also calculated two photon absorption (TPA) spectrum
using the MELDF code\cite{meld} at the SCI level. The right panel
of Fig. \ref{fig:EA} shows our calculated TPA spectrum for PPP-8.
The spectrum features an intense peak located at 3.04 eV, which exhibits
strong dipole coupling with the lowest one-photon $1B_{u}$ state,
and therefore is the $mA_{g}$ state\cite{mazumdar-mAg}, located
at 6.08 eV. Thus, our result for $mA_{g}$ state obtained from the
EA spectrum is in excellent agreement with that from the TPA.

\section{Conclusions and Future Directions}

\label{sec:conclusions}

In this paper we have described our Fortran 90 program which solves
the HF equations for both the closed- and open-shell molecular systems
using the semi-empirical Hückel and PPP models. To demonstrate the
features of our code, we presented results of numerous test calculations
on various molecular systems. The reason behind developing the present
program is twofold: (a) to develop a code in a modern language such
as Fortran 90 which can carry out dynamic array allocation, and thus,
free the user from specifying and changing array sizes, and (b) to
provide an open software which will be widely available to users which
they can use and modify as per their needs. One possible generalization
of this code will be to include the capabilities for calculations
on infinite systems in one- and higher dimensions to calculate the
band structure and related properties using the P-P-P model. Work
along those directions is continuing in our group, and results will
be published as and when they become available. 
\begin{ack}
Authors thank the Department of Science and Technology (DST), Government
of India, for providing financial support for this work under Grant
No. SR/S2/CMP-13/2006.\end{ack}

\end{document}